# Human-AI Interaction for User Safety in Social Matching Apps: Involving Marginalized Users in Design

Position Paper for "Artificially Intelligent Technology for the Margins" Workshop

Douglas Zytko
Oakland University
Rochester, MI, United States
zytko@oakland.edu

Nicholas Furlo
Oakland University
Rochester, MI, United States
nbfurlo@oakland.edu

Hanan Aljasim
Oakland University
Rochester, MI, United States
hkaljasi@oakland.edu

## ABSTRACT

In this position paper we intend to advocate for participatory design methods and mobile social matching apps as ripe contexts for exploring novel human-AI interactions that benefit marginalized groups. Mobile social matching apps like Tinder and Bumble use AI to introduce users to each other for rapid face-to-face meetings. These user discoveries and subsequent interactions pose disproportionate risk of sexual violence and other harms to marginalized user demographics, specifically women and the LGBTQIA+ community. We want to extend the role of AI in these apps to keep users safe while they interact with strangers across online and offline modalities. To do this, we are using participatory design methods to empower women and LGBTQIA+ individuals to envision future human-AI interactions that prioritize their safety during social matching app-use. In one study, stakeholders identifying as LGBTQIA+ or women are redesigning dating apps to mediate exchange of sexual consent and therefore mitigate sexual violence. In the other study, women are designing multi-purpose, opportunistic social matching apps that foreground women's safety.

## CCS CONCEPTS

•Human-centered computing ~ Collaborative and social computing ~ Empirical studies in collaborative and social computing

## 1 Introduction

Mobile social matching apps are an ideal context for studying human-AI interaction. One, they are a modern implementation of context-aware computing, which has been a focal point of HCI research into explainable AI [8,9]. Two, they are a ubiquitous example of the impact of AI on our social lives due to the popularity of apps such as Tinder and Bumble [2].

Mobile social matching apps leverage contextual data from mobile devices to compute user matches. Such apps facilitate discovery of near-countless individuals in one's geographic area and, in turn, rapid face-to-face encounters with strangers for reasons ranging from sex, friendship, marriage, and even employment. Despite the potential to augment social life, mobile social matching apps are associated with alarming rates of sexual violence [5–7,10–12,15] and harassment [2,4,14] that have disproportionately impacted women and the LGBTQIA+ community.

Our prior work has explored how design of the social matching app Tinder inadvertently perpetuates, or fails to stop, sexual violence against women and LGBTQIA+ users [17]. In our current work are we are looking to the future and how mobile social matching apps could better leverage AI, amongst other emerging technologies, to facilitate user safety and serve as scalable solutions to sexual violence.

In this workshop we intend to spark conversation about the utility of participatory design as a method for involving the marginalized into the creation of AI-infused technology. We hope to spark such discussion by sharing lessons learned about how to apply participatory design methods to human-AI interaction in a social matching context. We can also share emerging co-design ideas that exemplify novel future uses of AI for safety of marginalized demographics. These ideas could serve as the basis for more rigorous prototyping and user assessment and could lead to generalizable advances in human-centered AI.

Below we first introduce participatory design and briefly explain its advantages and challenges as a method for designing future human-AI interactions that benefit marginalized demographics. To highlight some of the value we can bring to workshop discussions, we then introduce two ongoing participatory design studies we are conducting and some methodological lessons learned that we can share with workshop attendees. One study explores how to design human-AI interaction to mediate sexual consent exchange in dating apps, and the other explores human-AI interaction for women's safety in multi-purpose opportunistic social matching apps.

## 2 Participatory Design: A Path Towards AI that Works for the Margins

Participatory design [13], or co-design, describes methods that directly incorporate end-users and other stakeholders into the design of technology. *Participation* has long been championed in HCI research, most notably Feminist HCI [3], because it represents an opportunity for marginalized populations to inform technology designs that acknowledge and work for them. Participatory design often takes the form of synchronous workshops or focus groups in which stakeholders interact with researchers to produce novel sketches, mockups, or verbal ideas for technologies. This direct, proactive involvement of anticipated end-users can be instrumental in shaping positive human-AI partnership and addressing modern barriers like explainability and trust [1,16].

Yet the method is not without its challenges. AI is a complex technology with several pop culture personifications that provide the public with distorted and misguided interpretations. For participatory design to work effectively co-designers must understand the true scope of "what" they are designing. Crucially, this understanding must be conveyed succinctly without making them feel inept and without unduly simplifying the (future) capabilities of AI.

## 3  Participatory Design Study 1: Designing Opportunistic Social Matching Apps to Foreground Women's Safety

Mobile matching apps facilitate rapid face-to-face encounters through user recommendation algorithms that leverage contextual information from users' mobile devices such as relative location. As amounts of contextual data continue to rise, we can expect these systems to one day be able to compute "just in time" recommendations for immediate social opportunities in the physical world. Potential for interpersonal harm will also be elevated because users will have less time to assess ephemeral opportunities and associated risk.

In this participatory design study, we are working with teams of woman-identifying students to produce risk-conscious designs of tomorrow's social matching apps. We chose women as our demographic of focus because they are disproportionately harmed through matching app-use and because we wanted participants to derive comfort from being around others that come from the same demographic and may have had similar experiences.

### 3.1  Methodological Lessons Learned

For the participatory design in this study we have been forming groups of 8-10 women, and each group attends four 1-hour Zoom sessions. This ongoing study has produced various insights into how to support co-design of AI-human interaction. Some include:

**Lesson 1: Introduce human-AI interaction with scenarios.** While our participants tend to readily understand modern matching apps such as Tinder and Bumble, it was initially a struggle to convey futuristic matching apps that can leverage AI for extremely relevant and immediate social encounters, particularly opportunities that go beyond dating. We have found user scenarios to be the best way to inform participants of this future use. The scenarios take the form of storyboard panels and we identify a particular panel in the storyboard that we want participants to generate designs for.

**Lesson 2: Build the relationship over time.** We made the decision to spread out the participatory design activities over four 1-hour sessions. By having recurrent meetings with the same participants, we are able to develop rapport with them and enable prolonged discussion of specific aspects of AI-human interaction in the context of matching apps. For example, in one session we focus only on how the co-designers would want to control the AI in futuristic social matching apps. In another we talk only about potential AI-infused features to be used during face-to-face encounters to keep women safe.

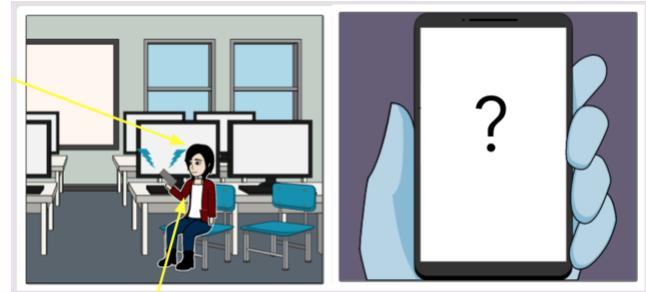

**Figure 1: By using storyboard-based scenarios we give co-designers a user to empathize with and a discrete opportunity for intervention with novel AI-infused tools.**

### 3.2  Emerging Findings

Our ongoing participatory design sessions have yielded several novel human-AI interaction possibilities for mobile matching apps of the future. One involves smart visibility, or the app's ability to proactively modify a user's visibility to others based on a series of context factors. For example, the app may set a woman's profile visibility as lower when in a geographic area she is unfamiliar with, or higher when social opportunities involve groups of people as opposed to individuals (safety in numbers). AI-driven discoverability can also be informed by individual users' "track records" on the app. Reporting or feedback features in the app could support a rich, multi-perspective corpus of feedback about individual users that can be used to identify problematic/risky users and generate understanding of users' personalities and compatibilities. The AI could leverage this feedback to compute user matches that are likely to be enjoyable and devoid of tension without the user having to actively fill out preference forms (e.g., matching two introverted users together, with that introversion being determined by past interactions).

## 4  Participatory Design Study 2: Mediating Consent Exchange through Dating Apps

Another of our ongoing studies explores how AI-infused design of mobile dating apps could mediate exchange of sexual consent and therefore mitigate sexual violence. This participatory design study involves LGBTQIA+ and women stakeholders because those demographics are at elevated risk of sexual violence [5–7,10–12,15].

This project is a follow-up to an interview study we conducted with users of various gender and sexual identities to pose explanations for the troubling association between dating app-use and sexual violence [17]. More specifically, the study explored how sexual consent exchange is mediated by the dating app Tinder. The study found that Tinder is informally viewed as a consent-exchange app, being used to infer and imply consent to sex without any verbal confirmation before making a sexual advance.

This current project considers how dating apps of the future could be intentionally designed as consent-exchange apps with augmentation from AI and other emerging technology such as mixed reality to prevent sexual violence.

### 4.1 Methodological Lessons Learned

Our participatory design sessions have taken the form of online, synchronous focus group sessions. They generally consist of 4-6 people and last about 2 hours long. This ongoing study has produced various insights into how to incorporate AI as a focal "material" in participatory design studies.

**Lesson 1: Clarify the preferred outcome of AI-infused technology.** During the first activity we ask each participant to provide an answer to the prompt: *"How do you think someone should give consent to a sexual encounter?"* We ask each participant to explain their response and discuss it with the group. This step is critical for clarifying what each participant expects future design intervention to achieve before we even introduce AI as a material for design, at which point participants may get caught up with designing cool new technology and lose sight of the original user goal that the technology was supposed to achieve.

**Lesson 2: Clarify the range of opportunities for intervention.** For the second activity we present participants with a timeline of someone using a dating application to find a positive consensual sexual experience. This timeline has three stages: the profile stage, the messaging stage, and the in-person meeting stage. We then split everyone into groups of two and assign each of these groups to design for a particular stage in the timeline. This is valuable in two regards. One, it helps mitigate redundancy in designs produced by the stakeholders. Two, we find that giving participants heavily structured prompts early on helps their brainstorming because they are less worried if they are designing correctly or otherwise doing what the researchers expect.

**Lesson 3: Not now, 10 years in the future.** We noticed early on in data collection that participants' ideas were largely derivative of existing dating app designs. Suspecting that they were too cautious to think outside the box, we now prompt participants to design an AI-infused dating app that would exist "10 years in the future." We also engage in a discussion before the AI design exercise begins that explores what AI means to the researchers and co-designers and stresses that all AI-infused ideas, regardless of how believable they may seem, are encouraged because they give us the best understanding of what effective, positive human-AI interaction would look like. This future-thinking prompt has significantly improved the novelty and creativity in co-designed solutions.

### 4.2 Emerging Findings

So far our participants have been able to generate a number of novel design ideas for human-AI interaction. One group imagined AI-infused map features. They envisioned dating apps being able to learn which public locations are safe meeting places for users from marginalized identity groups. This logistical information could inform matching algorithms so that users could discover others who they are not only compatible with, but capable of meeting safely for a face-to-face encounter. Other participants imagined a dating app's AI informing users of a messaging partner's prior harassment incidents.

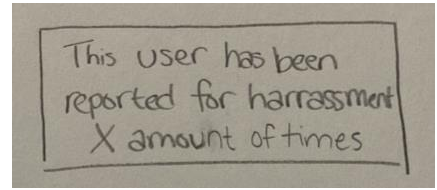

**Figure 2: Participants imagined a dating app's AI being used to detect and inform users of prior harassment, which could help them predict if an interaction partner may put them at risk of sexual violence.**

Another idea generated by participants was a user similarity detector, which provides insight for how AI outputs could be presented to users. Because members of the LGBTQIA+ community often use slang or coded language (e.g., "vanilla") to express their preferences and information about themselves, the AI would be able to detect and preserve such language when reporting user similarities. The similarity detector would also be able to identify dislikes or points of conflict that it knows a user values for predicting sexual violence risk; these would be outputted with particular colors for rapid digestibility and decision-making about a potential meeting partner.

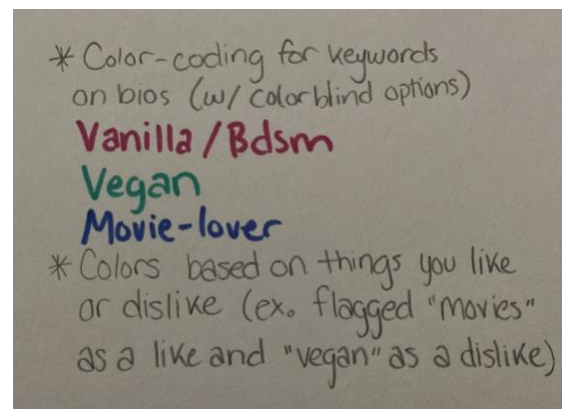

**Figure 3: Participants had suggestions on how an AI's output should be presented to users. Color-coding and retaining slang terminology were options posed for conveying key similarities and areas of conflict with potential meeting partners.**